\documentclass{article}

\PassOptionsToPackage{numbers, compress}{natbib}
\usepackage[preprint]{neurips_data_2024}

\usepackage[utf8]{inputenc} 
\usepackage[T1]{fontenc}    
\usepackage{hyperref}       
\usepackage{url}            
\usepackage{booktabs}       
\usepackage{amsfonts}       
\usepackage{nicefrac}       
\usepackage{microtype}      
\usepackage[svgnames]{xcolor}         
\usepackage{subcaption}
\usepackage{graphicx}
\usepackage{multirow}
\usepackage{pifont}
\usepackage{placeins}

\newcommand*\greencheck{\textcolor{ForestGreen}{\ding{52}}} 

\title{WebSuite: Systematically Evaluating \\ Why Web Agents Fail}

\author{%
  Eric Li \\
  Harvard University \\
  \texttt{ehli@college.harvard.edu}
  \And
  Jim Waldo \\
  Harvard University \\
  \texttt{waldo@g.harvard.edu}
}

\begin{document}

\maketitle

\begin{abstract}
  We describe WebSuite, the first diagnostic benchmark for generalist web agents, designed to systematically evaluate why agents fail. Advances in AI have led to the rise of numerous web agents that autonomously operate a browser to complete tasks. However, most existing benchmarks focus on strictly measuring whether an agent can or cannot complete a task, without giving insight on why. In this paper, we 1) develop a taxonomy of web actions to facilitate identifying common failure patterns, and 2) create an extensible benchmark suite to assess agents' performance on our taxonomized actions. This benchmark suite consists of both individual tasks, such as clicking a button, and end-to-end tasks, such as adding an item to a cart, and is designed such that any failure of a task can be attributed directly to a failure of a specific web action. We evaluate two popular generalist web agents, one text-based and one multimodal, and identify unique weaknesses for each agent. Because WebSuite can disaggregate task failures into specific action failures, this enables granular identification of which UX flows an individual agent has trouble with and immediately highlights promising avenues for improvement. These findings highlight the need for more focused benchmarking on where web agents go wrong to effectively improve agents beyond their weaker performance today.
\end{abstract}

\section{Introduction}

Recent developments in large language models (LLMs) have invigorated work in AI agents, including generalist web agents that use the browser as an interface to take actions on the web. In the future, much of web use might be via automated web agents that take actions on our behalf, which could increase productivity, expand free time, and generally augment human capabilities.

These generalist web agents are typically initialized with a natural language goal and starting URL. Then, they follow a process of parsing the HTML of a web page, using an LLM to decide the correct next action, such as clicking the search button, and repeating. \cite{deng2023mind2web, gur2023understanding, gur2024realworld, kim2023language, liu2023agentbench} Recent developments in multimodal model capabilities have also led to augmenting parsed HTML with screenshots of the webpage. \cite{zheng2024seeact} However, web agents are far from human performance, with the recent benchmark WebArena showing agents achieve only 14\% end-to-end (E2E) task success rate, compared to human performance of 78\%. \cite{zhou2024webarena}

We categorize existing work in web agent benchmarking in two types — those that evaluate performance on simpler, low-level tasks like clicking a button in MiniWoB \cite{shi2017wob} and those that evaluate performance on more complex, high-level tasks like purchasing an item in WebShop. \cite{yao2023webshop} While these benchmarks help us understand the (lack of) capabilities of these agents across many types of environments, why these agents fail is underexplored. To most effectively improve web agents, we need ways to diagnose their points of failure and identify the common patterns.

Thus, we introduce WebSuite, a diagnostic benchmark to systematically evaluate why web agents fail. We begin by leveraging digital literacy skills research and the popular Material Design web framework to create a taxonomy that breaks down any web operation into a high-level category, action, and specific interaction — providing multiple levels of detail for agent failure analysis. Then, we present an extensible task suite to evaluate web agents on these various web actions. Our task suite includes both individual tasks to directly evaluate action performance and E2E tasks to evaluate action performance as part of a longer, more strategic string of web actions. This gives a practical way for developers to begin evaluating and improving agents.

We evaluate two web agents on our benchmark, the popular open-source text-based agent natbot and the recent multimodal agent SeeAct. Because WebSuite is designed to attribute task failures to specific failures of our taxonomized actions, we are able to identify specific limitations for a given agent and report detailed success rates across a category, action, and interaction. For example, when evaluating natbot on individual tasks, we see it has difficulty filling forms, and when evaluating SeeAct on E2E tasks, we see it has difficulty clicking links in a search results list. Without WebSuite, identifying these failure patterns and how often they occur would require repeated manual observation of an agent as it completes various tasks.

While we present the first diagnostic benchmark for web agents, the concept of a diagnostic benchmark is not new. Prior works have proposed tools like CheckList for natural language processing and CLEVR and CARETS for visual question answering. \cite{jimenez2022carets, johnson2016clevr, ribeiro2020accuracy} Like CheckList, we draw inspiration from robust testing practices in software engineering and apply them to the burgeoning field of agentic AI systems. Like CLEVR, we believe that WebSuite can be used in conjunction with existing benchmarks to provide more comprehensive evaluation.

Ultimately, our findings indicate the need for more diagnostic tools to improve web agents and open new opportunities for further research. Our code is available at \texttt{\url{https://github.com/erichli1/websuite}}.

\section{Related Work}

\paragraph{Benchmarks for higher-level tasks.}
There are numerous benchmarks focused on evaluating generalist web agent performance on high-level tasks, involving following an overarching goal to take multiple actions. GAIA involves connecting numerous information-seeking actions together to get to a specific factual answer (\textit{e.g.}, How many studio albums were published by Mercedes Sosa between 2000 and 2009). \cite{mialon2023gaia} WebShop uses a simplified shopping environment loaded with scraped Amazon data, testing whether agents can purchase items (\textit{e.g.}, I am looking for x-large, red color women faux fur lined winter warm jacket coat, and price lower than 70.00 dollars). \cite{yao2023webshop} Mind2Web uses a series of snapshots of real websites with crowdsourced tasks, testing whether agents can accurately complete each step of a task in diverse settings (\textit{e.g.}, Find one-way flights from New York to Toronto). \cite{deng2023mind2web} WebArena uses open source versions of popular sites like Reddit, testing whether agents can complete E2E tasks (\textit{e.g.}, Upvote the newest post in deep learning subreddit). \cite{zhou2024webarena} There also exist benchmarks across computer and mobile environments, such as OSWorld, OmniACT, and Android in the Wild. \cite{kapoor2024omniact, rawles2023android, xie2024osworld} Generally, these higher-level evaluations are useful in assessing overall web agent capabilities but only provide a single signal of task success — did the agent complete the task or not — without answering the question of why. To identify where the agent is consistently going wrong, one would have to manually observe every action the agent takes in these environments.

\paragraph{Benchmarks for lower-level tasks.}
On the other side of the spectrum, there exist MiniWoB and MiniWoB++ as methods for evaluating the ability of agents to complete lower-level tasks, involving completing a simple goal with few steps. MiniWoB consists of a list of 100 tasks to test agents on specific capabilities in a controlled environment (\textit{e.g.}, click a specific link, drag numbers into sorted ascending order, use the terminal to delete a file). \cite{shi2017wob} MiniWoB++ extended these tasks to include more complex interaction patterns such as those requiring soft reasoning (\textit{e.g.}, select the words similar to "tragic") and stochastically varying layouts (\textit{e.g.}, the order of form fields changes for each test instantiation). \cite{liu2018reinforcement} That said, there exists a difference between being told directly to complete a low-level task (\textit{e.g.}, click the item labeled Next) and being able to complete this task when it is part of a long string of steps (\textit{e.g.}, "change privacy settings to Friends Only" which would involve navigating through a menu to go to the Privacy page).

WebSuite aims to bridge these two types of benchmarks and make it easy to identify common failure patterns through more contained individual tasks and see how success on these tasks might change in E2E instantiations.

\section{Taxonomy of Web Actions}

As discussed, when a web agent is trying to complete a task, it repeatedly outputs an action at each screen. Because each of these actions can either lead the agent towards or away from achieving the goal, we can use these intermediate actions as signal on the capabilities of an agent. Ultimately, any failure of an task can be traced to failures at specific actions. Thus, by taxonomizing the types of web actions, we can understand common modes of failure across multiple tasks.

\subsection{Categories and Actions}

We adapt Van Deursen and Dijk's framework of digital skills \cite{deursen2008measuringds} to generate categories and actions, as seen in Table~\ref{table:taxonomy}. We maintain similar high-level categories but modify the specific actions within each category to be more atomic, comprehensive, and relevant to modern web use. Operational actions involve interacting with the basic user experience (UX) patterns like a checkbox, link, or multi-select. Navigational actions involve interacting with navigational structures such as menus or forward/back buttons. Informational actions might involve retrieval like finding an answer on a web page, submission like filling out a form, or a combination of both like searching for a specific item.

We exclude their proposed "strategic skills" category because it represents a higher-level type of agent ability and can be decomposed into operational, navigational, and informational actions.

\begin{table}[hb]
    \caption{Taxonomy of web actions}
    \medskip
    \label{table:taxonomy}
    \centering
    \renewcommand{\arraystretch}{1.25}
    \begin{tabular}{ c | l }
        \toprule
        \textbf{Category} & \textbf{Action} \\
        \hline
         Operational & Click item \\ 
          & Type information \\  
          & Select option(s) \\
          \hline
         Navigational & Navigate to a URL \\ 
          & Navigate via a menu \\
          & Navigate forward/backward \\  
          \hline
         Informational & Find information on a page \\
          & Filter information on a page \\
          & Search \\
          & Fill out form \\
          & Review information for correctness \\
         \bottomrule
    \end{tabular}
\end{table}

This taxonomy should enable us to identify more specific methods of improvement for web agents. For example, if agents are bottlenecked by basic operational capabilities, it implies that there needs to be architecture improvements to enable agents to use the common UX patterns on the web. On the other hand, if agents are bottlenecked by informational capabilities, it implies that we need models with stronger reasoning capabilities. We can also step one level deeper and discuss aspects for improvement at the action-level. If agents are weak at filtering information but strong at other informational tasks, the problem might not be model reasoning ability but examples seen on how to specifically filter. If agents are unable to fill out forms, it may suggest that agents need better checking to see if the information they have filled is exactly correct. By breaking down web actions into this taxonomy, we enable the ability to improve agents in a targeted manner and compare agents across more fine-grained measures.

\subsection{Interactions}

To attain even more granular insights, we can break down each action into various types of interactions. For example, clicking a \texttt{link} is differentiable from clicking a \texttt{button}.

To create a (somewhat) comprehensive list of operational interactions, we leverage the Material Design 2 web components list\footnote{\texttt{\url{https://m2.material.io/components?platform=web}}} which helps us identify a number of UX patterns that are not included in MiniWoB or MiniWoB++, such as a \texttt{snackbar}, \texttt{switch}, \texttt{tooltip}, and \texttt{datagrid}.

While creating a list of interactions for each operational action means identifying a diverse list of web components, creating a list of interactions for navigational and informational actions is more about having diverse information represented across the same components. For example, there are a limited number of ways that a \texttt{menu} can be rendered but many more organizational hierarchies they can live under. See Appendix \ref{appendix:taxonomizedinteractions} for full list of interactions.

Unfortunately, not all interactions are mutually exclusive. For example, filling a \texttt{form} might rely on typing in multiple \texttt{text} fields, and filtering a \texttt{datagrid} might rely on both navigating a \texttt{menu} and using a \texttt{select}. This complication means that completing interactions composed of multiple more primitive interactions is inherently limited by the basic ability to complete those more primitive interactions.

\section{WebSuite}

Our goal is to create both individual tasks that can evaluate each interaction in the taxonomy and create E2E tasks that can evaluate interaction success rates across many steps. The benchmark environment uses a React frontend and Flask backend.

\subsection{Individual tasks}

\subsubsection{Creating tasks for each interaction}

We start with creating a logged UX component library that extends Material UI\footnote{\texttt{\url{https://mui.com/material-ui/}}} for the basic web components represented in the list of interactions. Each component in our library sends a log to the backend to be written to a designated text file. For example, the logged \texttt{text} component sends a log of what was typed with a 500 millisecond debounce.

\begin{figure}
    \centering
    \subcaptionbox{Make the volume louder}{\fbox{\includegraphics[width=.475\linewidth]{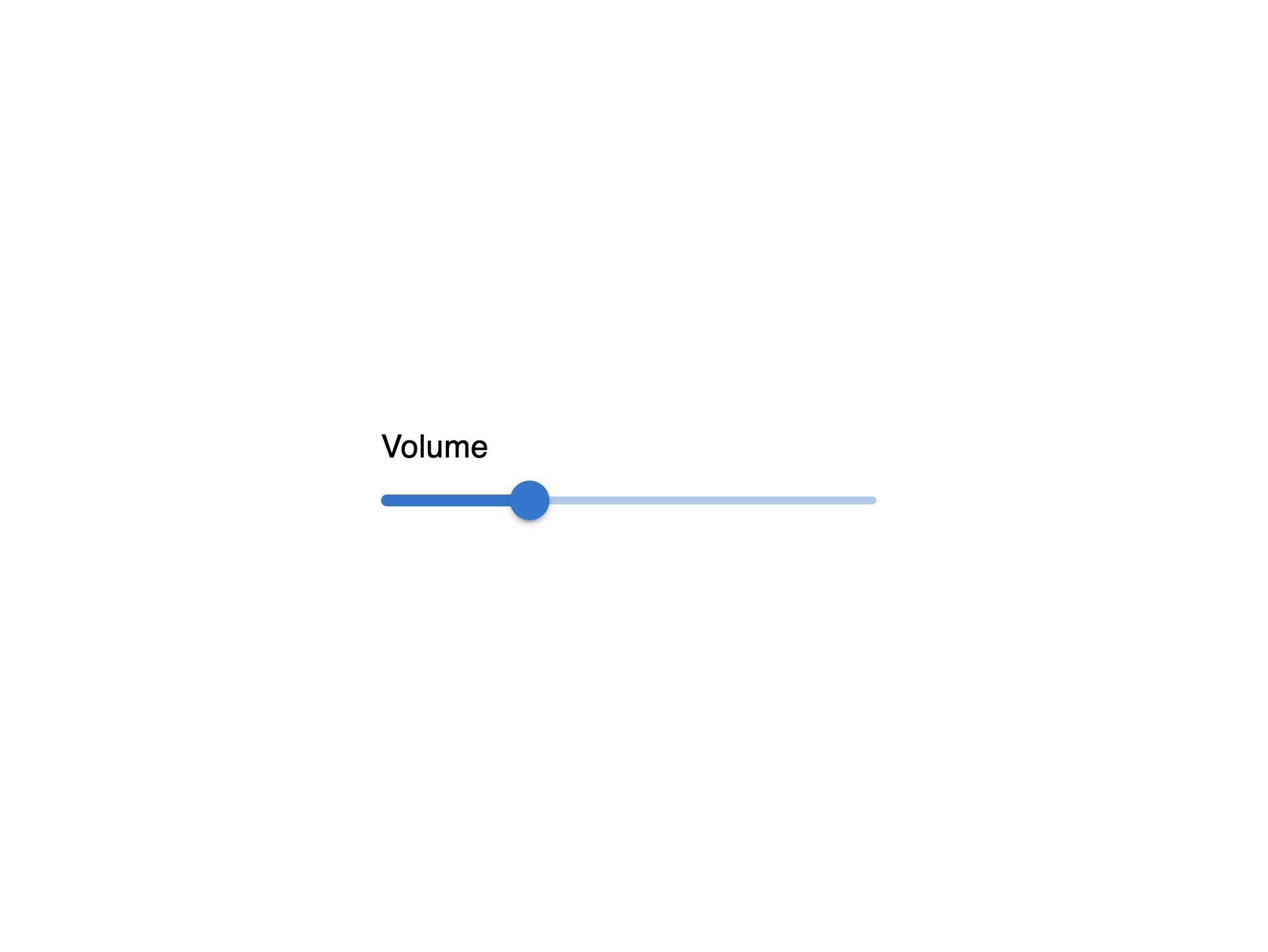}}}\hspace{.2em}%
    \subcaptionbox{Please filter for orders where the country is USA}{\fbox{\includegraphics[width=.475\linewidth]{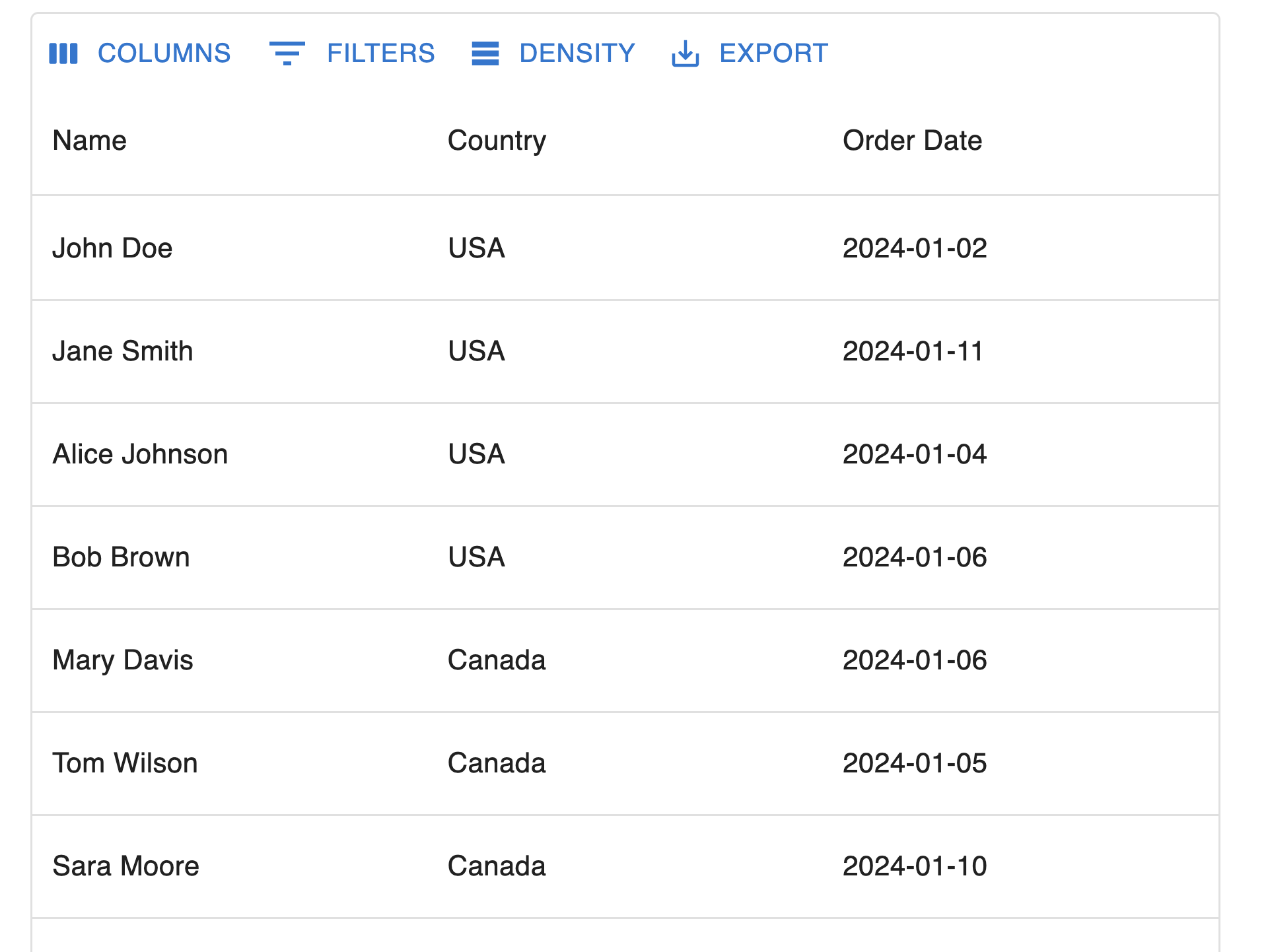}}}%
    \caption{Example individual tests}
    \label{fig:indexample}
\end{figure}

Individual tasks flow intuitively from each interaction. We implement tasks for a core subset of the interactions, and the code is designed to be easily extensible to add more tasks, or even more interactions, actions, and categories. Some interactions include multiple tasks to account for different uses. For instance, when shown a \texttt{switch}, one task evaluates whether the agent can turn on a \texttt{switch} from an off state while another task evaluates whether it can recognize a \texttt{switch} is already off and do nothing. The full list of interactions with implemented tasks is accessible in Appendix \ref{appendix:taxonomizedinteractions} and example screenshots of tasks are shown in Figure \ref{fig:indexample}.

\subsubsection{Evaluating agents}

Each task includes a starting path such as \texttt{/ind/click?test=button}, goal such as \textit{Turn on do not disturb}, and success criteria. Success criteria consists either of checking the content of the logs or evaluating the submitted material, such as for forms.

To add constraints to the individual tasks, we include a generous time limit (90 seconds for shorter tasks, 300 seconds for longer tasks) and an optional stopping condition, such as reaching a maximum length of two logs. We empirically found that agents often have trouble stopping, with initial tests ending up in dozens of clicks of a single element. Thus, by adding these constraints, we evaluate strictly whether an agent is capable of completing a task, not whether it completes the task and exits appropriately.

\subsection{E2E tasks}

\subsubsection{Creating the environment}

\begin{figure}[ht]
    \centering
    \subcaptionbox{Search results page}{\fbox{\includegraphics[width=.31\linewidth]{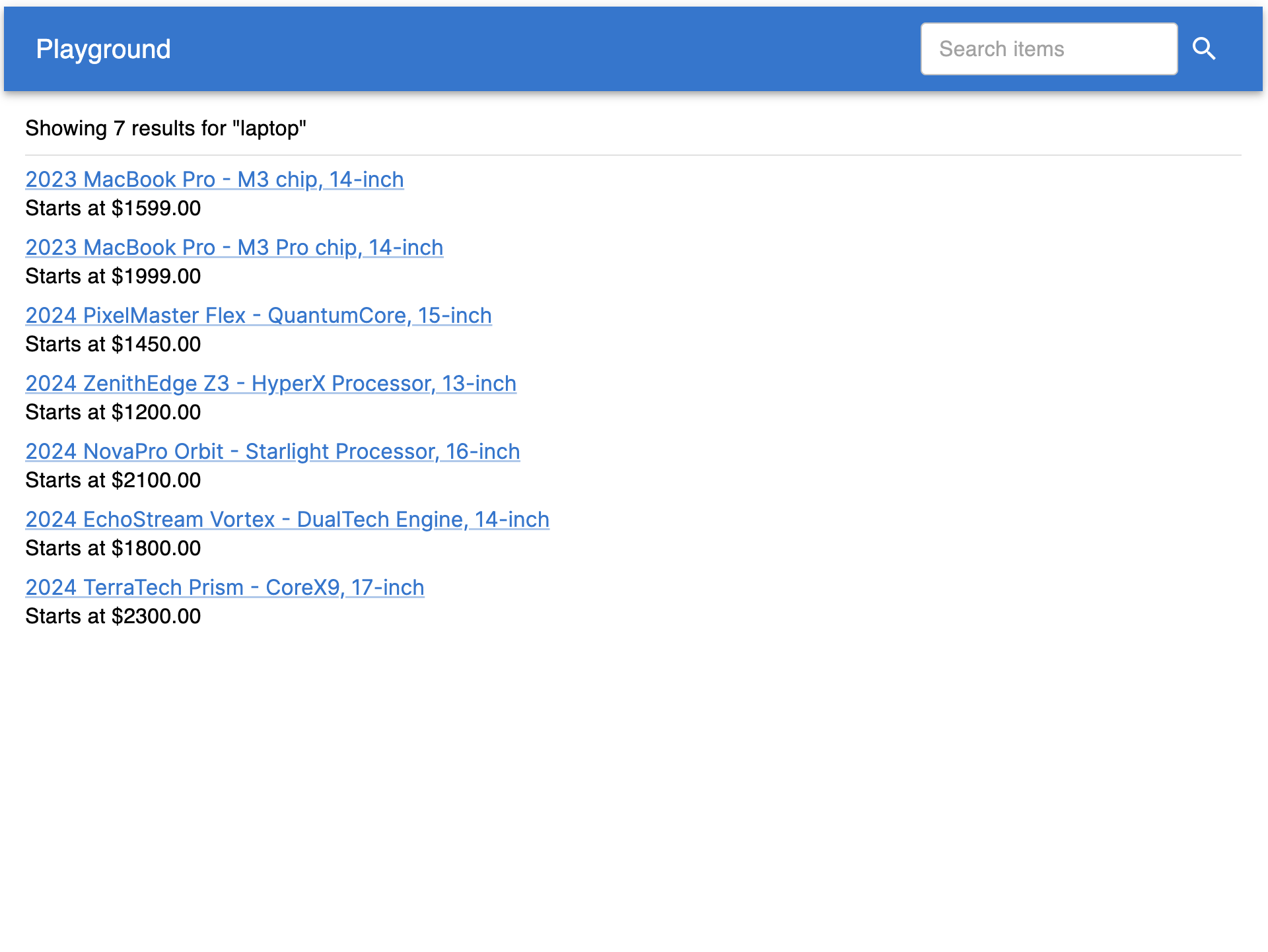}}}\hspace{.2em}%
    \subcaptionbox{Item page}{\fbox{\includegraphics[width=.31\linewidth]{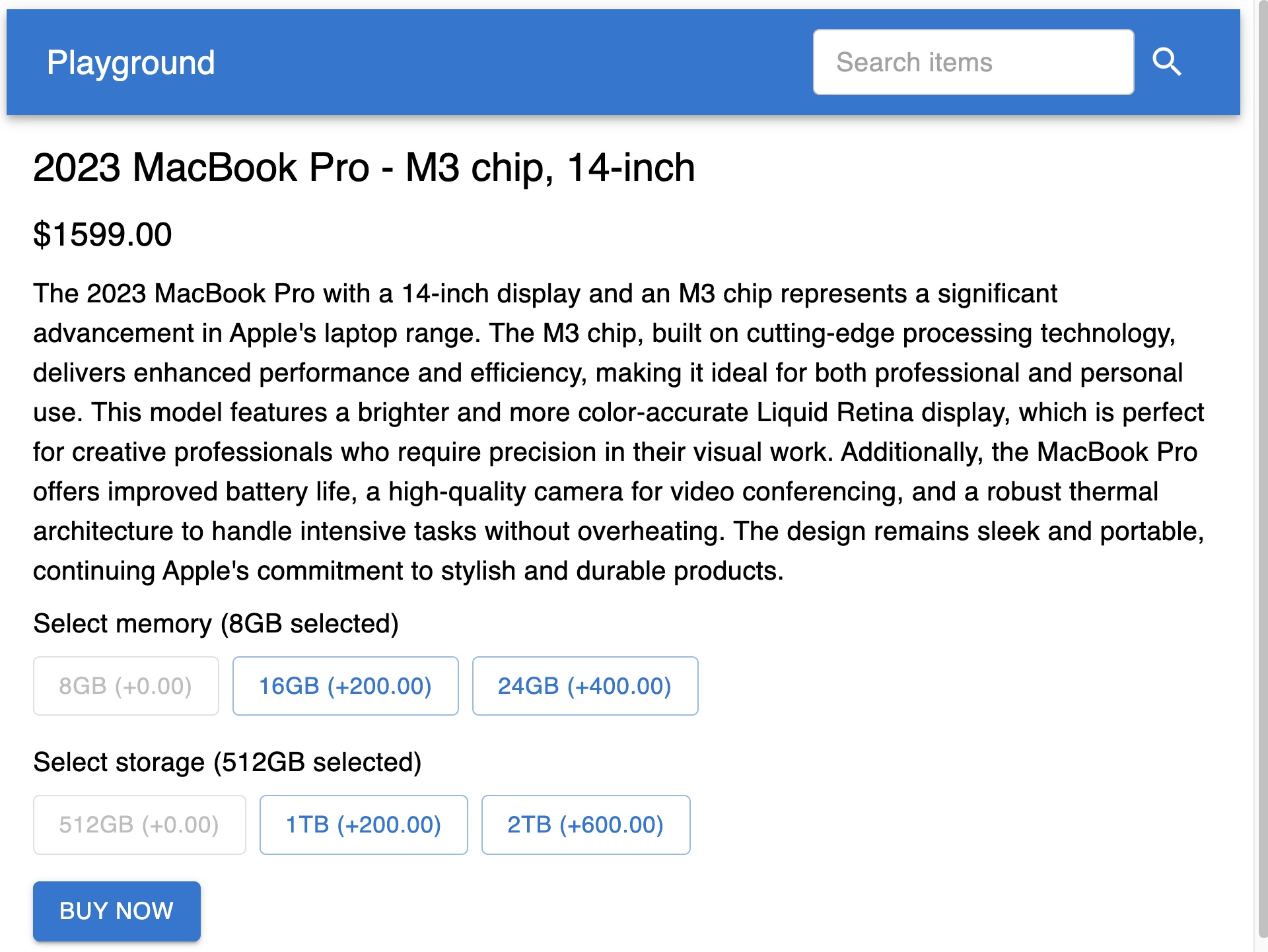}}}\hspace{.2em}%
    \subcaptionbox{Checkout page}{\fbox{\includegraphics[width=.31\linewidth]{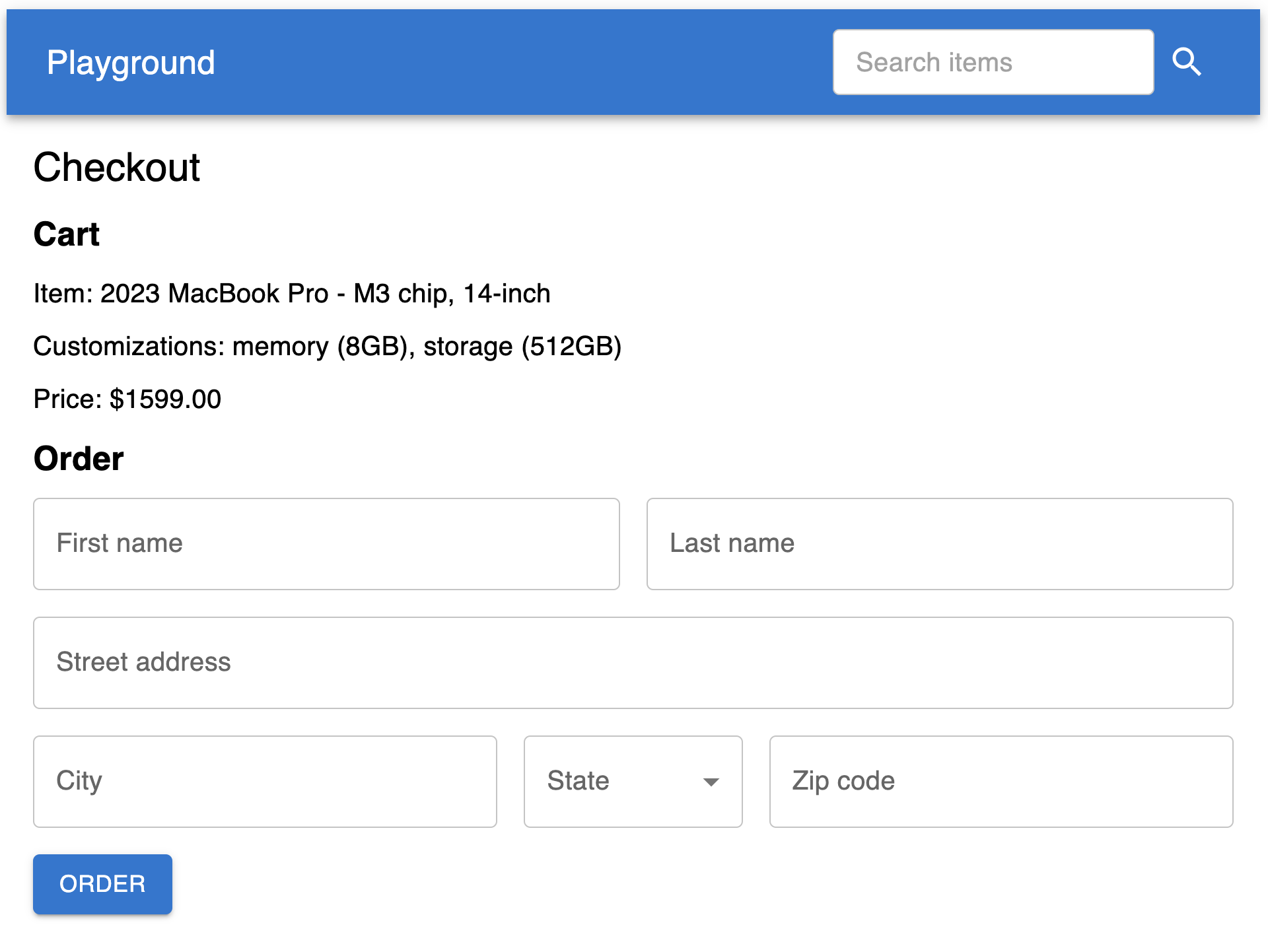}}}%
    \caption{Screenshots from shopping playground}
    \label{fig:e2eexample}
\end{figure}

Following prior work with WebShop and WebArena, we target creating a shopping site as a playground for web agent interaction. Using the logged UX components, we create a relatively barebones site, with search functionality, item page including buttons to select customizations, and a checkout page to fill a shipping address — see screenshots depicted in Figure \ref{fig:e2eexample}. Each time the agent interacts with the website, it generates a log, such as \texttt{click/iconbutton // Search} which indicates an \texttt{iconbutton} was clicked that had the label of "Search."

\paragraph{Checkpoints.} We design the shopping playground to include checkpoints along the way, in the form of URLs that fully encode any relevant previous states. For example, the search results page path is of the form \texttt{/search?query=<search terms>} indicating which terms were searched, and the checkout page path is of the form \texttt{/checkout?cart=<json of customized item>} indicating what item was purchased with which customizations.

\paragraph{Shopping items.} For items on the website, we draw inspiration from customizations in ordering a MacBook Pro\footnote{\texttt{\url{https://www.apple.com/shop/buy-mac/macbook-pro}}} and use ChatGPT to generate additional synthetic data following our format.

\paragraph{Tasks.} We begin with two relatively simple E2E tasks that include a number of our taxonomized interactions, and because the logged UX component library exists, it is easy to add more functionality and thus more tasks to this list. The two tasks are \texttt{order} and \texttt{add-to-cart}, and the goals are as follows:
\begin{enumerate}
    \item Please order a MacBook Pro M3 chip without additional customizations to be delivered to John Doe at 123 Main Street, Cambridge, MA 02138
    \item Please add a Macbook Pro with M3 Pro Chip to the cart with highest-tier customizations
\end{enumerate}

\subsubsection{Evaluating agents}

In defining an E2E task, we include a goal, a list of checkpoints and corresponding golden logs for each checkpoint, and an E2E success verifier. Golden logs specify what logs would occur in a fully correct string of steps at a specific checkpoint. For example, the search results page in Figure \ref{fig:e2eexample} would have one golden log — clicking the \texttt{link} of the correct result.

Because the URL fully encodes relevant previous states, we design the E2E success verifier to check whether the web agent navigated to the correct final page. As an example, the \texttt{order} verifier might check whether there exists a logged navigation to path \texttt{/thanks} and with query parameters of matching cart and shipping address. The verifier gives us the signal on E2E task completion, and we can compare the generated logs to the golden logs to evaluate exactly what types of web interactions in our taxonomy the agent is failing.

As mentioned, some interactions like \texttt{fill/basicform} are composed of more primitive operational interactions like \texttt{type/text}. Thus, we provide an option at the checkpoint-level to override what types of interactions are needed for this checkpoint. Without this option, we see very noisy failure patterns, such as many failed \texttt{type/text} interactions, when in reality, the agent has a broader inability to fill a form (\textit{e.g.}, it exits early after filling only one field).

To add constraints, for the full E2E evaluation, we give agents 300 seconds and automatically exit when they reach the final page of a task, such as the order confirmation page for the \texttt{order} task.

\section{Experiments}

\subsection{Setup}

We selected two agents to evaluate our benchmark and attain some preliminary results: natbot, a simple open-source web agent launched in 2022, \footnote{\texttt{\url{https://github.com/nat/natbot}}} and SeeAct, a recent multimodal web agent. \cite{zheng2024seeact} We designed the code to be easily extensible to testing and analyzing the results of more agents. We ran both agents on the full benchmark suite – both the individual and E2E tasks. We ran each agent on each task eight times to account for variability.

\subsection{Results} 

We display summary results for individual tasks in Table \ref{table:indresults}, and more detailed results are available in Appendix \ref{appendix:detailedindresults}. We display results for E2E tasks in Table \ref{table:e2eresults} and the patterns of interaction failures in E2E tasks in Table \ref{table:indtaske2eresults}.

\begin{table}[ht]
    \caption{Individual task success rate, aggregated by action.}
    \medskip
    \label{table:indresults}
    \centering
    \renewcommand{\arraystretch}{1.25}
    \begin{tabular}{ c | l | c | c | c }
        \toprule
        \textbf{Category} & \textbf{Action} & \textbf{Trials} & \textbf{natbot} & \textbf{SeeAct} \\
        \hline
         Operational & \textit{(combined)} & 128 & 85.16\% & 76.17\% \\ 
         \cline{2-5}
          & Click & 72 & 73.61\% & 72.22\% \\
          \cline{2-5}
          & Type & 24 & 100\% & 95.83\% \\  
          \cline{2-5}
          & Select & 32 & 100\% & 70.31\% \\
          \hline
         Navigational & Menu & 16 & 100\% & 95.83\% \\
          \hline
         Informational & \textit{(combined)} & 64 & 43.75\% & 40.64\% \\
         \cline{2-5}
          & Find & 32 & 53.13\% & 34.38\% \\
          \cline{2-5}
          & Filter & 16 & 50\% & 50\% \\
          \cline{2-5}
          & Fill & 16 & 18.75\% & 43.75\% \\
          \bottomrule
    \end{tabular}
\end{table}

\begin{table}[ht]
    \caption{E2E task completion rate. The checkpoints columns indicate successful completion rate for a specific checkpoint. The number in parentheses indicates the number of trials, which can be less than the starting number if the agent inconsistently makes it to certain checkpoints.}
    \medskip
    \label{table:e2eresults}
    \centering
    \renewcommand{\arraystretch}{1.25}
    \begin{tabular}{l | l | c | c  c  c  c}
        \toprule
        \textbf{Task} & \textbf{Agent} & \textbf{E2E} & \multicolumn{4}{c}{\textbf{Checkpoints}} \\
        \hline
        \multirow{3}{*}{order}
            & & & search & click item & add to cart & fill shipping \\
            \cline{4-7}
            & natbot & 12.5\% (8) & 100\% (8) & 100\% (8) & 100\% (8) & 12.5\% (8) \\
            & SeeAct & 0\% (8) & 100\% (8) & 0\% (8) & & \\
        \hline
        \multirow{3}{*}{add to cart}
            & & & search & click item & customize \\
            \cline{4-7}
            & natbot & 100\% (8) & 100\% (8) & 100\% (8) & 100\% (8) \\
            & SeeAct & 0\% (8) & 62.5\% (8) & 0\% (5) & \\
        \bottomrule
  \end{tabular}
\end{table}

\begin{table}[hbt]
    \caption{Interaction success rates during E2E tasks. The number in parentheses indicates the number of instances it encountered the interaction.}
    \medskip
    \label{table:indtaske2eresults}
    \centering
    \renewcommand{\arraystretch}{1.25}
    \begin{tabular}{ c | l | l | c | c }
        \toprule
        \textbf{Category} & \textbf{Action} & \textbf{Interaction} & \textbf{natbot} & \textbf{SeeAct} \\
        \hline
         Operational & \textit{(combined)} & & 100\% (56) & 64.44\% (45) \\ 
         \cline{2-5}
          & Click & \textit{(combined)} & 100\% (40) & 44.83\% (29) \\
          \cline{3-5}
          & & Button & 100\% (8) &  \\
          \cline{3-5}
          & & Icon button & 100\% (16) & 81.25\% (16) \\
          \cline{3-5}
          & & Link & 100\% (16) & 0\% (13) \\
          \cline{2-5}
          & Type & Text field & 100\% (16) & 100\% (16) \\
          \hline
         Informational & \textit{(combined)} & & 78.13\% (32)  & 0\% (13) \\
         \cline{2-5}
          & Fill & \textit{(combined)} & 56.25\% (16) &  \\
          \cline{3-5}
          & & Basic & 100\% (8) & \\
          \cline{3-5}
          & & Complex & 12.5\% (8) & \\
          \cline{2-5}
           & Search & Select result & 100\% (16) & 0\% (13) \\
          \bottomrule
    \end{tabular}
\end{table}

\FloatBarrier

\section{Discussion}

From our results, we can begin to draw more granular conclusions about the interactions that web agents fail and identify specific opportunities of improvement for individual agents.

\paragraph{Individual results.} At the category level, we see from Table \ref{table:indresults} that the tested agents are relatively capable operationally and navigationally but struggle with informational tasks. This is in line with other research that has highlighted weak performance on higher-level benchmarks like Mind2Web but strong performance on lower-level benchmarks like MiniWoB. \cite{zheng2024synapse} That said, there are still clear gaps in basic operational abilities, evident from the interaction-level statistics in Appendix \ref{appendix:detailedindresults}, such as interactions with a \texttt{slider}, \texttt{switch}, or \texttt{tooltip}, and filtering a \texttt{datagrid} or filling a \texttt{complex form}. While \texttt{slider} and \texttt{switch} inabilities can be explained by a basic lack of mouse abilities — dragging and hovering — the inability to filter a \texttt{datagrid} or fill a \texttt{complex form} is less intuitive given the fundamental operational capabilities exist. Focusing on natbot, this clues us to explore the datagrid filter task where it is instructed: \textit{Please filter for orders where the country is USA}. We find that natbot filters for names matching "USA" rather than for countries matching "USA", and thus, improving natbot might mean enhancing it with an LLM with stronger reasoning ability or even defining specific policies \cite{sodhi2024step} that handle the more complex workflow of filtering.

Looking at SeeAct's results, we see different areas for improvement. At the category level, we find it to be weaker operationally but have similar performance navigationally and informationally compared to natbot. Stepping one level deeper, we see it is entirely unable to complete a \texttt{select} interaction, which also contributes to its 0\% success rate at \texttt{fill/complexform}. This tells us that ensuring it can use \texttt{select} components would be an important first improvement step.

\paragraph{E2E results.} While the individual results are useful to find improvements, the E2E patterns of failure tell us about the more interesting higher-level tasks. For example, natbot's core problem is its inability to complete a \texttt{fill/complexform} interaction. It has perfect performance on all other tasks, so developers improving natbot should focus on its form-filling abilities.

On the other hand, with SeeAct, despite high performance of the \texttt{click/link} task in the individual evaluation, we see it is has 0\% success rate in the E2E environment — a big bottleneck to agent performance and (part of) the answer to why E2E task success is so low. Note that using WebSuite does not entirely replace the process of observing actual agent behavior but rather makes it easier to see patterns across many tasks. For example, only until observing SeeAct failing the \texttt{click/link} interaction do we see the underlying problem is it tries to click the larger container \texttt{<div>} rather than the actual \texttt{<a>} link.

\subsection{Looking forward}

\paragraph{Data analysis of failure patterns.} Because WebSuite moves beyond the broad success or failure signal towards fine-grained evaluations, we start getting more data than we can easily analyze, and thus, data anlysis could contribute even more detailed insights. For example, automatically surfacing which interaction failures contribute to which task failures becomes a data question. We discussed how the \texttt{click/link} task was weak for SeeAct; did this problem show up in the order or add to cart task? While this question is trivial in our two tasks scenario, these types of analyses become more meaningful as more E2E tasks get added.

\paragraph{Evaluate on diverse and real-world actions.} As discussed more thoroughly in Section \ref{section:limitationsandimpact}, many websites may not look like our WebSuite site. Thus, there exists a need to benchmark web actions on a more diverse and real-world environment. Extending this work could involve creating a detailed dataset of individual web interactions in our taxonomy, following similar processes as Mind2Web, snapshotting popular online sites. \cite{deng2023mind2web}

\paragraph{Augment existing benchmarks with taxonomy.} In its simplest form, each task in a benchmark could be tagged to include multiple types of actions, and success rates on the full task could be aggregated to identify failure patterns. In a more robust form, where there are specific steps like in the Mind2Web benchmark, one could tag each step with a corresponding web interaction in our taxonomy.

\paragraph{Automatically identify failures based on trajectories.} Pan et al. showcased how LLMs could be used to autonomously evaluate the trajectories of agent performance. \cite{pan2024autonomous} While they use "evaluator" LLMs to give a signal of task completion or failure, augmenting evaluators with our taxonomy could enable autonomously identifying the failure points of web agents. One could feed in a failed-task trajectory and our taxonomy to an evaluator and have it output what type of action it failed at. If performant, this would replace the need for extensive tagging of E2E benchmarks with relatively simple post-trajectory analysis.

\section{Limitations and Potential Societal Impact} \label{section:limitationsandimpact}

\paragraph{Specific to our web environment.} Our benchmark is rooted in the web design decisions we made. For example, before we added an aria-tag to the \texttt{iconbutton}, natbot was unable to interact with the \texttt{iconbutton} because the HTML lacked the necessary semantic information. While we took steps to mitigate this issue, such as by using the popular Material UI framework, there exists huge diversity in how real-world websites are constructed. Developers might use different design patterns, employ significant styling, and generally follow different web practices. Thus, our interaction, action, and even category success rates might not be transferable to different websites.

\paragraph{Simplified web environment.} Our benchmark also exists in a very simplified environment, with minimal CSS styling, besides the Material UI defaults, and pages that include little extraneous information to whatever task is being evaluated at hand. For example, clicking a single button may be a trivial task on a page with one button, but in a highly-stylized page, clicking the right button could prove difficult. As mentioned, we think there is lots of room to extend our taxonomy to more diverse and real-world environments.

\paragraph{Generalist web agents face dual-use concerns.} Our benchmark provides clear paths for agents to improve their capabilities. However, more capable technologies face dual-use concerns where they are both more powerful for helpful use cases like assisting users with monotonous desk work and harmful use cases like manipulation or scamming. We are excited about more discussions of the ethics of generalist agents, such as Gabriel et al.'s Ethics of Advanced AI Assistants.\cite{gabriel2024ethics}

\section{Conclusion}

We present a taxonomy of web actions and WebSuite, the first benchmark designed to evaluate web agent failure patterns across both lower-level and higher-level task instantiations. We performed a basic analysis of how two generalist web agents, natbot and SeeAct, perform and show how WebSuite can be used to quickly identify specific improvements for both agents. These efforts highlight the need for more fine-grained analysis of agent failure points and queue up meaningful opportunities for further work.

\bibliographystyle{abbrvnat}
\bibliography{bibliography}


\newpage
\appendix
\section{Appendix}

\subsection{Taxonomized interactions} \label{appendix:taxonomizedinteractions}

Full list of interactions for each category and action. \greencheck\ indicates at least one task has already been implemented for the interaction. Please see the code tfor more task details.

\begin{table}[ht]
    \caption{Operational actions and interactions}
    \medskip
    \centering
    \renewcommand{\arraystretch}{1.25}
    \begin{tabular}{ l l l }
        \toprule
        \textbf{Category} & \textbf{Action} & \textbf{Interaction} \\
        \hline
         Operational & Click item & \greencheck\  Accordion \\ 
          & & \greencheck\ Button \\
          & & \greencheck\ Dialog button \\
          & & \greencheck\ Dropdown menu \\
          & & \greencheck\ Icon button \\
          & & \greencheck\ Link \\
          & & \greencheck\ Slider \\
          & & \greencheck\ Snackbar \\
          & & \greencheck\ Switch \\
          
          & & Drawer \\
          & & Tab \\
          & & Floating action button \\
          \cline{2-3}
          & Type information & \greencheck\ Date \\  
          & & \greencheck\ Phone \\
          & & \greencheck\ Text field \\
          \cline{2-3}
          & Select option(s) & \greencheck\ Checkbox \\
          & & \greencheck\ Datagrid row \\
          & & \greencheck\ Multicheck \\
          & & \greencheck\ Select \\
          & & Radio \\
          & & Chips \\
         \bottomrule
    \end{tabular}
\end{table}

\begin{table}[ht]
    \caption{Navigational actions and interactions}
    \medskip
    \centering
    \renewcommand{\arraystretch}{1.25}
    \begin{tabular}{ l l l }
        \toprule
        \textbf{Category} & \textbf{Action} & \textbf{Interaction} \\
        \hline
         Navigational & Navigate to a URL & Navigate to arbitrary page \\ 
         \cline{2-3}
          & Navigate via a menu & \greencheck\ Basic menu \\
          & & \greencheck\ Nested menu \\
          \cline{2-3}
          & Navigate forward/backward & Navigate across arbitrary pages \\  
         \bottomrule
    \end{tabular}
\end{table}

\begin{table}[ht]
    \caption{Informational actions and interactions}
    \medskip
    \centering
    \renewcommand{\arraystretch}{1.25}
    \begin{tabular}{ l l l }
        \toprule
        \textbf{Category} & \textbf{Action} & \textbf{Interaction} \\
        \hline
         Informational & Find information on a page & \greencheck\ Accordion \\
         & & \greencheck\ Dialog button \\
         & & \greencheck\ Paragraphs \\
          & & \greencheck\ Tooltip \\
         & & Table \\
         \cline{2-3}
          & Filter information on a page & \greencheck\ Filter datagrid \\
          & & \greencheck\ Sort datagrid \\
          & & Filter search results \\
          & & Sort search results \\
          & & Filter spreadsheet \\
          & & Sort spreadsheet \\
         \cline{2-3}
          & Search & Write appropriate search query \\
          & & Select appropriate search result \\
          \cline{2-3}
          & Fill out form & \greencheck\ Basic form \\
          & & \greencheck\ Complex form \\
          \cline{2-3}
          & Review information for correctness & Confirm form is correct \\
          & & Change incorrect form elements \\
         \bottomrule
    \end{tabular}
\end{table}

\clearpage

\subsection{Detailed individual results} \label{appendix:detailedindresults}

\begin{table}[ht]
    \caption{Individual task success rates. For interactions with multiple tasks, we weight the trials by the number of tasks to avoid biasing aggregated action-level results towards those interactions. For example, there are four tasks for \texttt{slider} so each trial is weighted by a factor of one fourth during action-level aggregation. This ensures each interaction contributes equally to action performance.}
    \medskip
    \centering
    \renewcommand{\arraystretch}{1.25}
    \begin{tabular}{ c | l | l | c | c | c }
        \toprule
        \textbf{Category} & \textbf{Action} & \textbf{Interaction} & \textbf{Trials} & \textbf{natbot} & \textbf{SeeAct} \\
        \hline
         Operational & \textit{(combined)} & & 128 & 85.16\% & 76.17\% \\ 
         \cline{2-6}
          & Click & \textit{(combined)} & 72 & 73.61\% & 72.22\% \\
          \cline{3-6}
          & & Accordion & 8 & 100\% & 100\% \\
          \cline{3-6}
          & & Button & 8 & 100\% & 100\% \\
          \cline{3-6}
          & & Dialog button & 8 & 100\% & 87.5\% \\
          \cline{3-6}
          & & Dropdown menu & 8 & 100\% & 100\% \\
          \cline{3-6}
          & & Icon button & 8 & 100\% & 87.5\% \\
          \cline{3-6}
          & & Link & 8 & 100\% & 37.5\% \\
          \cline{3-6}
          & & Slider & 32 & 0\% & 0\% \\
          \cline{3-6}
          & & Snackbar & 8 & 12.5\% & 100\% \\
          \cline{3-6}
          & & Switch & 8 & 50\% & 37.5\% \\
          \cline{2-6}
          & Type & \textit{(combined)} & 24 & 100\% & 95.83\% \\
          \cline{3-6}
          & & Date & 8 & 100\% & 87.5\% \\
          \cline{3-6}
          & & Phone & 8 & 100\% & 100\% \\
          \cline{3-6}
          & & Text field & 8 & 100\% & 100\% \\
          \cline{2-6}
          & Select & \textit{(combined)} & 32 & 100\% & 70.31\% \\
          \cline{3-6}
          & & Checkbox & 8 & 100\% & 100\% \\
          \cline{3-6}
          & & Grid row & 8 & 100\% & 93.75\% \\
          \cline{3-6}
          & & Multicheck & 8 & 100\% & 87.5\% \\
          \cline{3-6}
          & & Select & 8 & 100\% & 0\% \\
          \hline
         Navigational & Menu & \textit{(combined)} & 16 & 93.75\% & 81.25\% \\
          \cline{3-6}
          & & Basic & 8 & 100\% & 100\% \\
          \cline{3-6}
          & & Nested & 8 & 87.5\% & 62.5\% \\
          \hline
         Informational & \textit{(combined)} & & 64 & 43.75\%  & 40.63\% \\
         \cline{2-6}
          & Find & \textit{(combined)} & 32 & 53.13\% & 34.38\% \\
          \cline{3-6}
          & & Accordion & 8 & 100\% & 50\% \\
          \cline{3-6}
          & & Dialog button & 8 & 12.5\% & 87.5\% \\
          \cline{3-6}
          & & Paragraphs & 8 & 100\% & 0\% \\
          \cline{3-6}
          & & Tooltip & 8 & 0\% & 0\% \\
          \cline{2-6}
          & Filter & \textit{(combined)} & 16 & 50\% & 50\% \\
          \cline{3-6}
          & & Filter datagrid & 8 & 0\% & 0\% \\
          \cline{3-6}
          & & Sort datagrid & 8 & 100\% & 100\% \\
          \cline{2-6}
          & Fill & \textit{(combined)} & 16 & 18.75\% & 43.75\% \\
          \cline{3-6}
          & & Basic & $8$ & $25\% (\pm 30)$ & $87.5\% (\pm 23)$ \\
          \cline{3-6}
          & & Complex & 8 & $12.5\% (\pm 23)$ & $0\% (\pm 0)$ \\
          \bottomrule
    \end{tabular}
\end{table}

\end{document}